\documentclass[prl,twocolumn,showpacs,superscriptaddress,a4paper]{revtex4}
\usepackage{graphicx}
\usepackage{psfrag}
\usepackage{amsmath}
\usepackage{amssymb}
\usepackage{amsfonts}

\begin{document}
%
%
\title{Fluctuations in  Stationary non Equilibrium States}


\author{L. Bertini}
\affiliation{Dipartimento di Matematica, Universit\`a di Roma
``La Sapienza", Piazza A. Moro 2, 00185 Roma, Italy}

\author{A. De Sole}
\affiliation{Department of Mathematics, MIT,
77 Massachusetts Avenue, Cambridge, MA 02139-4307, USA }

\author{D. Gabrielli}
\affiliation{Dipartimento di Matematica, Universit\`a dell'Aquila,
67100 Coppito, L'Aquila, Italy}

\author{G. Jona--Lasinio} 
\affiliation{Dipartimento di Fisica,
Universit\`a di Roma ``La Sapienza", Piazza A. Moro 2, 00185 Roma,
Italy}

\author{C. Landim} 
\affiliation{IMPA, Estrada Dona Castorina 110, J. Botanico, 22460 Rio
de Janeiro, Brasil, \\
 and CNRS UPRES-A 6085, Universit\'e de Rouen,
76128 Mont-Saint-Aignan Cedex, France}


\begin{abstract}
In this paper we formulate a dynamical fluctuation theory for
stationary non equilibrium states (SNS) which covers situations in a
nonlinear hydrodynamic regime and is verified explicitly in
stochastic models of interacting particles.  In our theory a crucial
role is played by the time reversed dynamics.  Our results include the
modification of the Onsager--Machlup theory in the SNS, a general
Hamilton--Jacobi equation for the macroscopic entropy and a non
equilibrium, non linear fluctuation dissipation relation valid for a
wide class of systems.
\end{abstract}

\pacs{05.20.-y, 05.40.-a, 05.60.-k}

\maketitle
%
%

%
%
%
\narrowtext

The Boltzmann--Einstein theory of equilibrium thermodynamic
fluctuations, as described for example in Landau--Lifshitz \cite{LL},
states that the probability for a fluctuation from equilibrium in a
macroscopic region of volume $V$ is proportional to $ \exp\{V\Delta S
/ k\}$ where $\Delta S$ is the variation of entropy density calculated
along a reversible transformation creating the fluctuation and $k$ is
the Boltzmann constant. This theory is well established and has
received a rigorous mathematical formulation in classical equilibrium
statistical mechanics via the so called large deviation theory
\cite{La}.  The rigorous study of large deviations has been extended
to hydrodynamic evolutions of stochastic interacting particle systems
\cite{KOV}. In a dynamical setting one may asks new questions, for
example what is the most probable trajectory followed by the system in
the spontaneous emergence of a fluctuation or in its relaxation to
equilibrium.  The Onsager--Machlup approach \cite{ON2} answers
precisely to this question: in the situation of a linear hydrodynamic
equation, that is, close to equilibrium, the most probable emergence
and relaxation trajectories are one the time reversal of the
other. Developing the methods of \cite{KOV}, this theory has been
extended to nonlinear regimes \cite{JLV}.  Onsager--Machlup assume the
reversibility of the microscopic dynamics; however microscopically non
reversible models were constructed where the above results still hold,
\cite{GJL1,GJL2}.

Emergence of large fluctuations, including Onsager--Machlup symmetry,
has been observed in stochastically perturbed gradient type electronic
devices \cite{LM}.  In their work, these authors study also non
gradient (i.e.\ non reversible) systems and observe violation of
Onsager--Machlup symmetry.

In the present paper we formulate a general theory of large deviations
for irreversible processes, i.e.\ when detailed balance condition does
not hold. This question was previously addressed in \cite{EY}.
Natural examples are boundary driven stationary non equilibrium states
(SNS), e.g.\ a thermodynamic system in contact with two reservoirs,
but our theory covers, as a special case, also the model systems
considered in \cite{LM}.  In our approach a crucial role is played by
the time reversed dynamics with respect to the stationary non
equilibrium ensemble.

Our results are:

1. The Onsager--Machlup relationship has to be modified: the emergence
of a fluctuation takes place along a trajectory which is determined by
the time reversed process. 

2.  We show that the macroscopic entropy solves a
Hamilton--Jacobi equation generalizing to a thermodynamic context
known results for finite dimensional Langevin equations \cite{FW} as
those studied in \cite{LM}. The Hamilton--Jacobi equation can be
solved perturbatively if we consider not too large fluctuations.

3. We test our theory in a stochastic model of interacting particles
system, the boundary driven zero range process, in which we perform
the computations explicitly.  In particular it is possible to
construct the microscopic time reversed process and to write the
macroscopic entropy in a closed form.

4. For a large class of systems we obtain a non equilibrium non
linear fluctuation dissipation relationship which links the
macroscopic evolution of the system and of its time reversal to the
thermodynamic force.

\medskip

We are interested in many body system in the limit of infinitely many
degrees of freedom. The basic assumptions of our theory are the following.

1) The microscopic evolution is given by a Markov process $X_t$ which
represents the configuration of the system at time $t$. 
This hypothesis probably is not so restrictive because also the
Hamiltonian case discussed in \cite{EPR} in the end is reduced to the
analysis of a Markov process. The stationary non equilibrium state
(SNS) is described by a stationary, i.e.\ invariant with respect to
time shifts, probability distribution $P_{st}$ over the trajectories of
$X_t$.

2) The macroscopic behavior of the system is described by
diffusion type hydrodynamical equations of the form
\begin{equation}
\partial_t \rho = 
\sum^d_{i,j}\partial_{u_i}\{D_{i,j}(\rho)\partial_{u_j} \rho\}
= F(\rho)
\label{H}
\end{equation}
where $\rho=\rho_t(u)$ represents in general a vector of thermodynamic
variables, e.g.\ the densities of different species of particles, and
$D_{i,j}$ is a matrix acting on this vector. The interaction with the
reservoirs appears as boundary condition to be imposed on solutions of
(\ref{H}).  We assume that there exists a unique stationary solution
$\bar\rho$ of (\ref{H}), i.e.\ a profile $\bar\rho(u)$ which satisfies
the appropriate boundary conditions such that $F(\bar\rho)=0$.

These equations are derived from the underlying microscopic dynamics
through an appropriate scaling limit.  The hydrodynamic equations 
represent laws of large numbers with respect the
probability measure $P_{st}$ conditioned on an initial state
$X_0$. The initial conditions for (\ref{H}) are determined by
$X_0$. Of course many microscopic configurations give rise to the same
value of $\rho_0(u)$. In general $\rho_t(u)$ is an appropriate limit
of a $\rho_N(X_t)$ as the number $N$ of degrees of freedom diverges.

3) The stationary measure $P_{st}$ admits a principle of large deviations
describing the fluctuations of the thermodynamic variables appearing
in the hydrodynamic equations. This means the following. 
The probability that in a macroscopic volume $V$ containing
$N$ particles  the evolution of the variable $\rho_N$
deviates from the solutions of the hydrodynamic equations and is close
to some trajectory ${\hat{\rho}}_t$, is exponentially
small and of the form
\begin{eqnarray}
\label{LD}
P_{st}\left( 
\rho_N(X_t) \approx \hat{\rho}_t(u), t\in [t_1, t_2]\right) 
&\approx& e^{-N[S(\hat \rho_{t_1}) + J(\hat{\rho})] } 
\nonumber \\
&=& e^{-N I(\hat{\rho})}
\end{eqnarray}
where $J(\hat{\rho})$ is a functional which vanishes if
${\hat{\rho}}_t$ is a solution of (\ref{H}) and $S(\hat \rho_{t_1})$
is the entropy cost to produce the initial value ${\hat{\rho}}_{t_1}$.
We adopt the convention for the entropy sign opposite to the usual
one, so that it takes the minimum value in the equilibrium state. We
also normalize it so that $S(\bar\rho)=0$.  Therefore $J(\hat{\rho})$
represents the extra cost necessary in order that the system follow
the trajectory ${\hat{\rho}}_t$. Finally $\rho_N(X_t) \approx
\hat{\rho}_t(u)$ means closeness in some metric.  This formula is a
generalization of the Boltzmann--Einstein formula in which we set the
Boltzmann constant $k=1$.

4) Let us denote by $\theta$ the time inversion operator defined by
$\theta X_t = X_{-t} = X^*_t$. The probability measure $P^*_{st}$
describing the evolution of the time reversed process $ X_{t}^*$ is
given by the composition of $P_{st}$ and ${\theta}^{-1}$ that is
\begin{equation}
\label{P*}
P^*_{st} (X_t^*=\phi_t, t\in [t_1,t_2]) = 
P_{st} (X_t=\phi_{-t}, t\in [-t_2,-t_1])
\end{equation}
We assume that the time reversed SNS also admits a hydrodynamic
level of description. 

Let $L$ be the generator of the microscopic dynamics.  We remind that
$L$ induces the evolution of functions of the process according to the
equation $\partial_t E[f(X_t)] = E[(Lf)(X_t)]$, where $E$ stands for
the expectation with respect to $P_{st}$ conditioned on the initial
state $X_0$ \cite{FE}.  The time reversed dynamics is generated by the
adjoint $L^*$ of $L$ with respect to the invariant measure $P_{in}$,
that is $E^{P_{in}}[fLg]= E^{P_{in}}[(L^*f)g]$.The measure $P_{in}$,
which is the same for both processes, is a distribution over the
configurations of the system and formally satisfies
$L^*P_{in}=0$. $E^{P_{in}}$ is the expectation with respect to
$P_{in}$ and $f,g$ are functions of the configuration.  We require
that also the evolution generated by $L^*$ admits a hydrodynamic
description, that we call the adjoint hydrodynamics, which, however,
is not necessarily of the same form as (\ref{H}).  In fact the adjoint
hydrodynamics can be non local in space.

In order to avoid confusion we emphasize that what it is usually called 
an equilibrium state, as distinguished from a SNS, corresponds to the
special case $L^* = L$, i.e.\ the detailed balance principle holds.  
In such a case $P_{st}$ is invariant under time reversal and 
the two hydrodynamics coincide.

\medskip

We now derive a fundamental consequence of our assumptions,
that is the relationship between the action functionals
$I$ and $I^*$ associated to the dynamics $L$ and $L^*$.
 From equation (\ref{P*}) and our assumptions it follows that also 
$P^*_{st}$ admits a large deviation principle with functional 
$I^*$ given by
\begin{equation}
I^*_{[t_1,t_2]}(\hat \rho) = I_{[-t_2,-t_1]}(\theta \hat \rho)
\label{I} 
\end{equation}
with obvious notations. More explicitly this equation reads
\begin{equation}
S({\hat \rho}_{t_1}) + J^*_{[t_1,t_2]}(\hat \rho) =
S({\hat \rho}_{t_2}) + J_{[-t_2,-t_1]}(\theta \hat \rho) 
\label{J}
\end{equation}
where ${\hat \rho}_{t_1}, {\hat \rho}_{t_2}$ are the initial and final
points of the trajectory and $S({\hat \rho}_{t_i})$ the entropies
associated with the creation of the fluctuations $\hat {\rho}_{t_i}$
starting from the SNS. 
The functional $J^*$ vanishes on the solutions of the adjoint
hydrodynamics. To compute $J^*$ it is necessary to know the
macroscopic entropy $S$, which is determined by the microscopic
invariant measure $P_{in}$.  We shall see that $S$ can be also
obtained as the solution of a Hamilton--Jacobi equation involving
only macroscopic quantities.  From (\ref{J}) we can already obtain the
generalization of Onsager-Machlup relationship for SNS.

The physical situation we are considering is the following.
The system is macroscopically in the stationary state $\bar\rho$
at $t=-\infty$ but at $t=0$ we find it in the state ${\hat \rho}_0$. 
We want to determine the most probable trajectory followed in the
spontaneous creation of this fluctuation. According to (\ref{LD}) this
trajectory is the one that minimizes $J$ among all trajectories
connecting $\bar\rho$ to ${\hat \rho}_0$ in the time interval
$[-\infty,0]$. From (\ref{J}) we have
\begin{equation}
J_{[-\infty,0]}(\hat \rho) = J^*_{[0,\infty]}(\theta \hat \rho) +
S({\hat \rho}_0)
\label{OM}
\end{equation}
The right hand side is minimal if $J^*_{[0,\infty]}(\theta \hat
\rho)=0$ that is if $\theta \hat \rho$ is a solution of the adjoint
hydrodynamics.  The existence of such a relaxational solution is due
to the fact that the stationary solution $\bar\rho$ is attractive also
for the adjoint hydrodynamics.  We have therefore the following
generalization of Onsager-Machlup to SNS

{\sl{``In a SNS the spontaneous emergence of a macroscopic fluctuation
takes place most likely following a trajectory which is the time
reversal of the relaxation path according to the adjoint hydrodynamics"}}

\medskip
 
Let us assume  $J$ of the form 
\begin{equation}
J_{[t_1,t_2]}(\hat \rho)={\frac {1}{2}}
 \int_{t_1}^{t_2} dt \left\langle W , K(\hat\rho) W \right\rangle 
\label{J2}
\end{equation} 
where $W = \partial_t {\hat\rho} - F(\hat\rho)$, $F(\hat\rho)$ has
been defined in (\ref{H}), $\langle \cdot,\cdot\rangle$ denotes
integration in the space variable and $K(\hat\rho)$ is an appropriate
positive kernel which we assume to be known and reflects, at the
macroscopic level, the stochasticity of the system.  This form of $J$
is typical for diffusion processes described by finite dimensional
Langevin equations \cite{FW} and is what we expect if the hydrodynamic
equations are of the diffusion type (\ref{H}). This is also what we
find for the model discussed later.

From (\ref{J}) or (\ref{OM}) we have that the entropy is related to
$J$ by
\begin{equation} 
\label{ff1}
S(\rho)= \inf_{\hat \rho}  J_{[-\infty,0]}(\hat \rho)
\end{equation}
where the infimum is taken over all trajectories 
connecting $\bar\rho$ to $\rho$. 
Therefore $S$ must satisfy the Hamilton--Jacobi equation
associated to the action functional $J$.  A simple calculation gives
\begin{equation}
{\frac {1}{2}}
\left\langle {\frac {\delta S}{\delta \rho}},
K^{-1}(\rho){\frac {\delta S}{\delta \rho}}\right\rangle
+\left\langle{\frac {\delta S}{\delta \rho}},F \right\rangle = 0
\label{HJ}
\end{equation}
One can try to solve this functional derivative equation by successive
approximations. Let $\bar\rho(u)$ be the stationary profile in the
SNS.  If we expand $S$ as a Volterra functional series in the argument
$\rho-\bar\rho$, equation (\ref{HJ}) reduces to a system of equations
for the kernels of the expansion which can be solved by iteration.  
Applications of (\ref{HJ}) to variuos models will be discussed 
in \cite{NOI}.

\medskip
We consider now the so called zero--range process which models
a nonlinear diffusion of a lattice gas, see e.g.\ \cite{KL}.
The model is described
by a positive integer variable $\eta_{\tau}(x)$ representing the number
of particles at site $x$ and time $\tau$ of a finite 
lattice which for simplicity we assume one--dimensional. 
The particles jump with rates $g(\eta (x))$   
to one of the nearest-neighbor sites $x+1, x-1$ with probability $1/2$.
The function $g(k)$ is non decreasing and $g(0)=0$. We assume that our
system interacts with two reservoirs of particles in positions $N$ and
$-N$ with rates $p_{+}$ and $p_{-}$ respectively. The microscopic
dynamics is then defined by the generator \cite{DF}
\begin{eqnarray}
\label{GEN}
(L_N f)(\eta) &=&  {\frac {1}{2}}\sum_{x=-N}^N g(\eta (x))[
\nabla^{x,x+1} f + \nabla^{x,x-1}f] \\
&+&  {\frac {1}{2}}p_{+} [f(\eta^N) - f(\eta)] + 
{\frac {1}{2}}p_{-} [f(\eta^{-N}) - f(\eta)] 
\nonumber
\end{eqnarray}
where $\nabla^{x,x\pm 1}f = f(\eta^{x,x\pm 1}) -f (\eta)$ and
$$
\eta^{x,y}(z) \; =\; \left\{
\begin{array}{ccl}
\eta(z) &\hbox{if}& z\neq x,y \\
\eta(z) -1 &\hbox{if}& z=x \\
\eta(z) +1  &\hbox{if}& z=y \; ,
\end{array}
\right.
$$
that is, $\eta^{x,y}$ is the configuration obtained from $\eta$ when a
particle jumps from $x$ to $y$. Similarly $\eta^{N,N+1} (N) = \eta (N)
- 1$, $\eta^{-N,-N-1} (-N) = \eta (-N) - 1$ and $\eta^N (\eta^{-N})$
is the configuration $\eta$ after addition of a particle at the point
$N (-N)$.

It is remarkable that the invariant measure  for this process can be
constructed explicitly \cite{DF}. It is the gran--canonical 
measure obtained by the product of the marginal distributions 
\begin{equation}
P_N(\eta(x) = k) = {\frac {\lambda_N^k (x)}{g(1)\cdots g(k)}} Z_N^{-1}(x)
\label{INV}
\end{equation}
where
\begin{equation}
\lambda_N (x) = {\frac {p_+ - p_-}{2(N+1)}}x + {\frac {p_+ + p_-}{2}}
\end{equation}
and
\begin{equation}
\label{Z=}
Z_N (x) = 1 + \sum_{k=1}^{\infty}{\frac {\lambda_N^k (x)}{g(1)
\cdots g(k)}}
\end{equation}

We emphasize that, if $p_-\neq p_+$, the generator $L_N$ is not
self--adjoint with respect to the invariant measure so that the
process is different from its time reversal and detailed balance does
not hold.

Let us introduce now the macroscopic time $t= \tau /N^2$ and space 
$u = x/N$ and the empirical density
\begin{equation}
\rho_N(t,u) = {\frac {1}{N}}\sum_{x=-N}^N\eta_{N^2 t }(x) 
\delta (u - x/N)
\label{ED}
\end{equation}
Then one can prove that in the limit $N\rightarrow \infty$ the
empirical density (\ref{ED}) tends in probability to a continuous
function $\rho_t(u)$ which satisfies the following hydrodynamic
equation
\begin{equation}
\partial_{t} \rho = {\frac {1}{2}} 
\partial_u \{ D(\rho) \partial_u \rho \}
= {\frac {1}{2}}\partial_u^2 \phi (\rho)=F(\rho)
\label{HZ}
\end{equation}
where  $\phi(\rho)$ is the inverse function of
\begin{equation}
\label{rfi}
\rho (\phi) = \frac {1}{ Z(\phi)} \sum_{k=0}^\infty 
\frac {\phi^k}{g(1)\cdots g(k)} \, k
\end{equation}
where $Z(\phi)$ is the normalization constant as in (\ref{Z=}).  The
function $\phi$ is well defined because the right hand side of
(\ref{rfi}) is strictly increasing in $\phi\ge 0$.  Then $\rho(\phi)$
is the equilibrium density corresponding to the activity $\phi$.  The
boundary conditions for (\ref{HZ}) are $\phi(\rho_{t}(\pm 1)) =
p_{\pm}$.

 From the knowledge of the invariant measure one can calculate the
adjoint generator $L^*_N$ given by
\begin{eqnarray}
\label{GEN*}
&& (L_N^* f)(\eta) = {\frac {1}{2}}\sum_{x=-N}^N g(\eta (x)) 
\frac{\lambda_N(x+1)}{\lambda_N(x)} \nabla^{x,x+1} f \\
&& \qquad\qquad + {\frac {1}{2}}\sum_{x=-N}^N g(\eta (x)) 
\frac{\lambda_N(x-1)}{\lambda_N(x)} \nabla^{x,x-1} f
\nonumber \\
&& + {\frac {1}{2}} \lambda_N(N)  [f(\eta^N) - f(\eta)] + 
{\frac {1}{2}} \lambda_N(-N)  [f(\eta^{-N}) - f(\eta)] \; .
\nonumber
\end{eqnarray}
Notice that the form of (\ref{GEN*}) is the same as (\ref{GEN}) with
the rates modified in such a way to invert the particle flux. From
(\ref{GEN*}) one can derive the adjoint hydrodynamics
\begin{equation}
\partial_{t} \rho  = {\frac {1}{2}}\partial_u^2 \phi (\rho)
- \alpha \partial_u \{{\frac {\phi (\rho)}{\lambda(u)}}\}=F^*(\rho)
\label{AHZ}
\end{equation}
with $\lambda (u) = {\frac {p_{+} - p_{-}}{2}}u + {\frac {p_{+} + p_{-}}{2}}$
and $\alpha = {\frac {p_{+} - p_{-}}{2}}$.
The boundary condition for (\ref{AHZ}) are the same as for (\ref{HZ}).
For the detailed computations we refer the reader to \cite{NOI}.

The second term on the right hand side of (\ref{AHZ}) is produced by
the new rates in (\ref{GEN*}).  As expected, it is proportional to the
difference of the chemical potentials.

In this model one can compute the action functionals $J(\hat\rho)$ and 
$J^*(\hat\rho)$. The result is 
\begin{eqnarray}
J_{[t_1,t_2]}(\hat\rho) &= & 
{\frac {1}{2}}\int_{t_1}^{t_2} \!dt\, 
\left\langle \nabla_u^{-1}W ,{\frac {1}{\phi(\hat\rho)}}
\nabla_u^{-1}W \right\rangle
\label{J1}
\\
J_{[t_1,t_2]}^*(\hat\rho)&=&
{\frac {1}{2}}\int_{t_1}^{t_2} \!dt \, 
\left\langle \nabla_u^{-1} W^*,
{\frac {1}{\phi(\hat\rho)}} \nabla_u^{-1} W^*
\right\rangle
\label{J*}
\end{eqnarray}
where, $W^* = \partial_t {\hat\rho} - F^*(\hat\rho)$.  Moreover $J$
and $J^*$ are defined to be infinite if $\hat\rho_t$ does not satisfy
the boundary conditions stated above.

The entropy $S(\rho)$ can be easily computed from the expression
(\ref{INV}) for the invariant measure
 \begin{equation}
S(\rho)=
\int \!du \: \left[ \rho(u) \log \frac {\phi(\rho(u))}{\lambda(u)}  
- \log \frac{Z(\phi(\rho(u)))}{Z(\lambda(u))}
\right]
\label{E}
\end{equation}

 From equations (\ref{J1}), (\ref{J*}) and (\ref{E}), one can 
verify explicitly equation (\ref{J}) and the generalized
Onsager-Machlup relationship.

\medskip

We deduce now a twofold generalization of the celebrated
fluctuation--dissipation relationship: it is valid in non equilibrium
states and in non linear regimes.
Such a relationship will hold provided the rate function 
$J^*_{[t_1,t_2]}$ of the time reversed process is of the form 
(\ref{J2}) with the same metric $K(\rho)$ but a different vector field
$F^*$ describing the hydrodynamic of the adjoint process, namely
\begin{equation}
J_{[t_1,t_2]}^*(\hat\rho)={\frac {1}{2}}\int_{t_1}^{t_2} \!dt\:
 \left\langle W^* ,K(\hat\rho) W^* \right\rangle
\label{J2*}
\end{equation}
By taking the variation of the equation (\ref{J}), a simple
computation gives 
\begin{equation}
\label{gfd}
F(\rho) +F^*(\rho) = - K(\rho)^{-1} \frac {\delta S}{\delta \rho}
\end{equation}
This relation holds for the non--equilibrium zero--range model
discussed before.  We also note that it holds for the equilibrium
reversible models for which the large deviation principle has been
rigorously proven such as the simple exclusion process \cite{KOV} and
the Landau--Ginzburg model \cite{DV} and its non--gradient version
\cite{Q}.  It is also easy to check that the linearization of
(\ref{gfd}) around the stationary profile $\bar \rho$ yields a
fluctuation--dissipation relationship which in equilibrium reduces to
the usual one.


\medskip

We are grateful to E. Presutti for insisting that the results obtained
for special models should actually exemplify general principles.  We
also thank A. De Masi, P. Ferrari, G. Giacomin and M. E. Vares for
very illuminating discussions.  D. G. was supported by FAPESP
98/11899-2 at IME-USP, Brazil and is very grateful to A. Galves and
P. Ferrari for the warm hospitality. He wishes to thank P. Markowich
and the University of Vienna for support during the final stage of the
work through the Wittgenstein award. L. B. and G. J-L. acknowledge the
support of Cofinanziamento MURST.

%
%

%
%

\end{document}